# Data-Driven Modeling of Landau Damping by Fourier Neural Operator

Shichen Wei
Key Lab for Information Science of Electromagnetic Waves (MoE)
Fudan University
Shanghai, P. R. China
21210720242@m.fudan.edu.cn

Yuhong Liu
Key Lab for Information Science of Electromagnetic Waves (MoE)
Fudan University
Shanghai, P. R. China
anifacel@163.com

Haiyang Fu*
Key Lab for Information Science of Electromagnetic Waves (MoE)
Fudan University
Shanghai, P. R. China
haiyang_fu@fudan.edu.cn

Chuanfei Dong*
Department of Astronomy
Boston University
Boston, Massachusetts, USA
dcfy@bu.edu

Liang Wang
Department of Astrophysical Sciences and
Princeton Plasma Physics Laboratory
Princeton University
Princeton, New Jersey, USA
liang.wang@princeton.edu

*Abstract*—The development of machine learning techniques enables us to construct surrogate models from data of direct numerical simulations, which has important implications for modeling complex physical systems. In this paper, based on the output from 1D Vlasov-Ampere simulations, we adopt the Fourier Neural Operator (FNO) to build surrogate models of Landau fluid closure for multi-moment fluid equations from kinetic simulation data. The trained FNO is able to obtain the heat flux using electron density as input, in agreement with the true value from kinetic simulations. We compare the physical quantities obtained using the FNO and Multilayer Perceptron (MLP) architectures, and found that the results of FNO are significantly better than that of MLP.

*Keywords—Landau Fluid Closure; FNO; MLP*

## I. INTRODUCTION

Plasma behaviors can be described by kinetic approaches or fluid models. Kinetic approaches are essentially statistical descriptions of the spatial-temporal evolution of distribution functions which are usually adopted to study the plasma behaviors on microscopic scales, while fluid models are continuum descriptions of macroscopic physical quantities. Since the computational costs of kinetic approaches are generally expensive, they are usually not affordable for large-scale and multi-scale problems [1-4]. People tend to solve fluid equations to reduce computational costs while capturing kinetic features of the plasma behaviors to some extent [1-3]. When deriving the fluid equations from velocity moments of the kinetic Vlasov equation, the higher-order moment quantities are always included in the lower-order moment equations. In order to close the system of equations (i.e., the fluid closure problem), it is necessary to use the lower-order moments to approximate the higher-order terms.

Due to the rapid development of neural network architectures, deep learning has emerged as a new approach to the scientific discovery of partial differential equations and to build surrogate models [5,7-11]. Early attempts have been made by, e.g., Ma et al. [5], who built surrogate models of Hammett-Perkins closure [6] with the multilayer perceptron (MLP), the convolutional neural network (CNN), and the discrete Fourier transform (DFT) network. Raissi et al. [12] proposed Physics-Informed Neural Networks (PINNs), which combine the neural network with the physical properties of the problem rather than deriving a solution based on data alone. PINNs can solve partial differential equations through sparse data and complete inversion of equation coefficients. It is noteworthy that Qin et al. [13] applied PINNs to construct a multi-moment fluid model with an implicit fluid closure included in the neural network that can capture Landau damping successfully. More recently, Li et al. [14] introduced the Fourier Neural Operator (FNO), a novel deep-learning architecture capable of learning the mapping between infinite-dimensional spaces of functions. The integral operator is restricted to a convolution and instantiated through a linear transformation in the Fourier domain.

This paper aims to construct an FNO network to build surrogate models of Landau fluid closure for multi-moment fluid models that can capture Landau damping from multiple sets of kinetic simulation data. We also compare the results with that of adopting a traditional ANN network. The paper is organized as follows: Section II shows the simulation results of Landau damping and describes the methods using the FNO and MLP architectures. Section III presents the predictions of the heat flux using the two neural network architectures, and Section IV gives the summary.

## II. PHYSICAL MODEL AND METHODOLOGY

### A. Physical Model

We study Landau damping of Langmuir waves by evolving the plasma phase space distribution function. The initial setup consists of an immobile, neutralizing ion background and multiple perturbation modes applied to the electron density.

$$n_e(x, t=0) = n_0(1 + A_1 \cos(k_1 x + \varphi_1) + \cdots + A_n \cos(k_n x + \varphi_n)) \quad (1)$$

$$n_i(x, t=0) = n_0 \quad (2)$$

where $n_0$ is the initial density of each species, $k_n$ is the wavenumber of the n[th] mode, $A_n$ the random amplitude, and $\varphi_n$ the random phase.

We obtain electron fluid quantities by taking velocity moments via integration of the distribution function f(x, v, t), including number density n(x, t), velocity u(x, t), pressure p(x, t), and heat flux q(x, t).

$$n(x,t) = \int f(x,v,t) dv \quad (3)$$

$$u(x,t) = \frac{1}{n(x,t)} \int v f(x,v,t) dv \quad (4)$$

$$p(x,t) = m_e \int (v-u)^2 f(x,v,t) dv \quad (5)$$

$$q(x,t) = m_e \int (v-u)^3 f(x,v,t) dv \quad (6)$$

$$T(x,t) = \frac{p(x,t)}{n(x,t)} \quad (7)$$

The Vlasov-Ampére equation is solved using Gkeyll [15]. The temperature T(x,t) is calculated from Eq. (7).

*B. Fourier Neural Operator*

Figure 1a shows the neural network architecture for the FNO network. The input a(x) is mapped to a higher dimensional space using a fully connected layer (FC1). The transformed feature is passed through four Fourier layers (FL). Finally, a fully connected layer (FC2) is used to obtain the final output u(x) with the desired dimensions. Figure 1b shows the architecture of a Fourier layer. The input goes through two paths in the Fourier layer. In the top path, the input undergoes a Fourier transform, linear transform, and inverse Fourier transform. In the bottom path, the input undergoes a linear transform W. The outputs from each path are summed together and undergo the ReLU activation function $\sigma$.

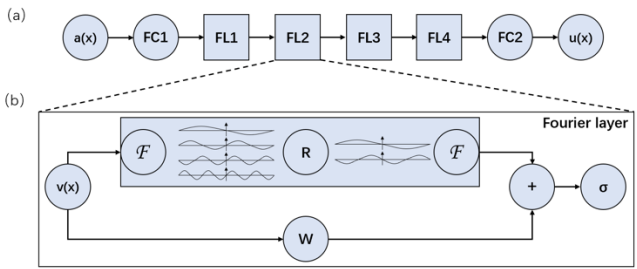

Fig.1. Neural network architecture of the FNO network.

*C. Multilayer Perceptron*

The neural network architecture of a multilayer perceptron (MLP) is shown in Figure 2. The neural network consists of one input layer, two hidden layers, and one output layer with 256 input and output variables, where each hidden layer has 64 neurons.

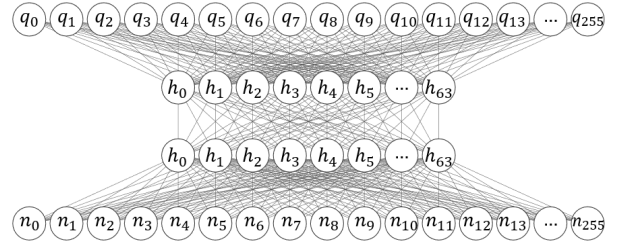

Fig.2. Neural network architecture schematics of MLP.

### III. MODELING RESULTS AND ANALYSIS

*A. Modeling Results Using the FNO Architecture*

We obtain 1000 sets of physical quantities with different initial conditions using Gkeyll, including density n(x, t) and heat flow q(x, t). We take the data from 0 to 10 $\omega_{pe}^{-1}$ and use all the points as one dataset. For the 1000 datasets, 80% of the data are used as the training set, 10% as the verification set, and 10% as the test set. Electron density n(x, t) and heat flux q(x, t) are shown in Figure 3.

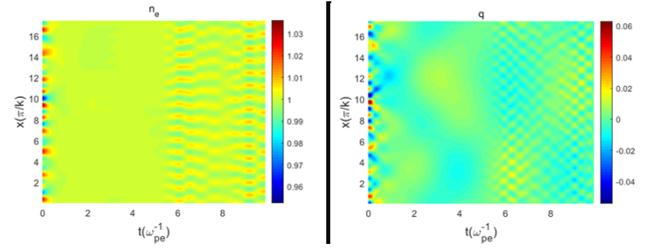

Fig.3. Spatial-temporal evolution of electron density and heat flux from one of the kinetic simulations.

Based on our 1X1V kinetic simulations, we construct neural networks to build surrogate models of Landau fluid closure for multi-moment fluid models from kinetic simulation data shown in Figure 3.

When tested on heat flux $q$, the heat flux $q$ produced by the FNO network and Gkeyll is remarkably similar and visually nearly identical (Figure 4). It indicates that the FNO network can capture an accurate fluid closure for multi-moment fluid models. One popular loss function for the regression problem is the relative error, which is defined in Eq. (8). For the test data set, the relative error of the FNO network is 10.8%, which shows that the FNO network can predict the heat flux $q$ reasonably well.

$$L = \frac{1}{N} \sum_{i=1}^{N} \frac{\left\| y_{true}^i - y_{pred}^i \right\|}{\left\| y_{true}^i \right\|} \quad (8)$$

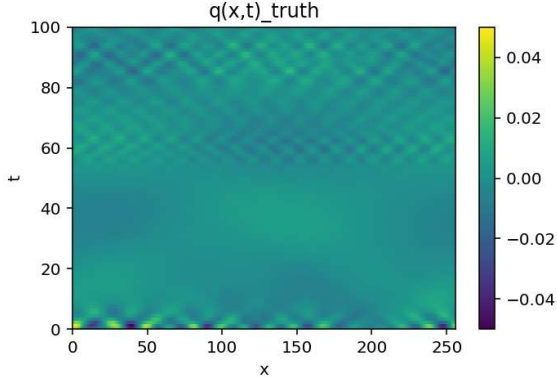

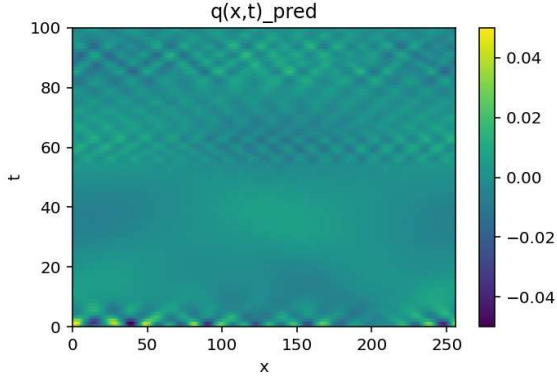

Fig.4. Comparison of the calculated heat flux, q, between Gkeyll and FNO.

### B. Modeling Results Using the MLP Architecture

We feed the same data into the MLP architecture, that is, 80% of the data is used for the training set, 10% of the data is used for the validation set, and 10% of the data is used for the test set. For the test data set, the relative error of the MLP is 84.0%. The predicted heat flux results by the MLP architecture are shown in Figure 5.

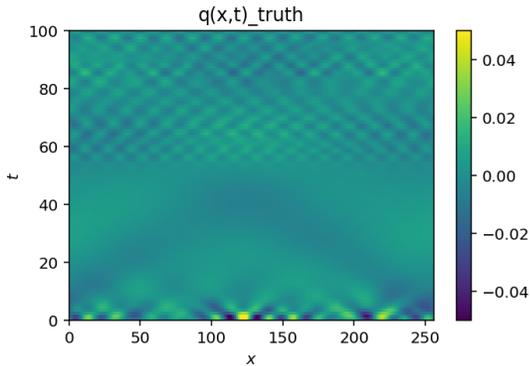

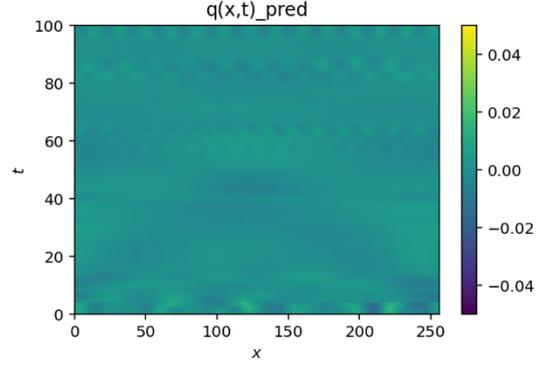

Fig.5. Comparison of the calculated heat flux, q, between Gkeyll and MLP.

| Neural Networks | FNO | MLP |
|---|---|---|
| Relative Error | 10.8% | 84.0% |

Table 1. Relative error by adopting FNO and MLP.

In summary, we present the spatial-temporal evolution of the heat flux obtained using the FNO and MLP architectures, as well as their comparison with the true values from kinetic simulations. Table 1 shows the relative errors by adopting FNO and MLP. It is clear that the physical quantities obtained by the surrogate model constructed by the FNO architecture are more accurate than that from the MLP architecture, mainly because FNO can be used to describe mappings between infinite-dimensional spaces and between different finite-dimensional approximations of those spaces.

## IV. CONCLUSION AND DISCUSSION

In this paper, we build a surrogate model of Landau fluid closure that can capture the collisionless Landau damping of electrostatic waves based on the FNO and MLP architectures. The obtained surrogate model can be integrated into multi-moment fluid models to close the system of equations. Compared with the MLP architecture, we found that the heat flux obtained from the FNO network agrees better with the true values from kinetic simulations. In addition, it is also possible to build such surrogate models using data from laboratory and spacecraft measurements that are capable of predicting other quantities. We plan to optimize the network further and aim to make it capable of modeling high-dimensional data with noise.

### ACKNOWLEDGMENTS

The authors from Fudan University were supported by the National Key Research and Development Program of China (2021YFA0717300) and China National Science Foundation (42074189). C.D. was supported by the U.S. Department of Energy under contract number DE-AC02-09CH11466.